\pdfoutput=1
\documentclass[twocolumn,aps,superscriptaddress,prl]{revtex4-1}

\usepackage[dvipsnames]{xcolor}
\usepackage{xspace}
\usepackage{times}
\usepackage{graphicx}
\usepackage{float}
\usepackage{dsfont}
\usepackage{amsmath}
\usepackage{stmaryrd}
\usepackage{natbib}
\usepackage[unicode=true,
  linktocpage,
  linkbordercolor={0.5 0.5 1},
  citebordercolor={0.5 1 0.5},
  linkcolor=blue]{hyperref}

\newcommand{\degr}[1]{\ensuremath{#1^\circ}\xspace}

\newcommand{\Bs}{\ensuremath{B^{\star}}\xspace}
\newcommand{\pd}{{\phantom{\dagger}}}

\begin{document}

\title{Uncovering Weyl Fermions in the Quantum Limit of NbP}

\author{K.~A.~Modic}
\affiliation{Max-Planck-Institute for Chemical Physics of Solids, Dresden, Germany 01187.}
\author{Tobias~Meng}
\affiliation{Institut f\"ur Theoretische Physik, Technische Universit\"at Dresden,
01062 Dresden, Germany.}
\author{Filip~Ronning}
\author{Eric~D.~Bauer}
\affiliation{Los Alamos National Laboratory, Los Alamos, NM, USA 87545.}
\author{Philip~J.~W.~Moll}
\affiliation{Max-Planck-Institute for Chemical Physics of Solids, Dresden, Germany 01187.}
\author{B.~J.~Ramshaw}
\affiliation{Laboratory of Atomic and Solid State Physics, Cornell University, Ithaca, NY, 14853.}

\date{\today}

\maketitle

% Abstract

\textbf{The Fermi surface topology of a Weyl semimetal (WSM) depends strongly on the position of the chemical potential. If it resides close to the band touching points (Weyl nodes), as it does in TaAs, separate Fermi surfaces of opposite chirality emerge, leading to novel phenomena such as the chiral magnetic effect. If the chemical potential lies too far from the nodes, however, the chiral Fermi surfaces merge into a single large Fermi surface with no net chirality. This is realized in the WSM NbP, where the Weyl nodes lie far below the Fermi energy and where the transport properties in low magnetic fields show no evidence of chiral Fermi surfaces. Here we show that the behavior of NbP in \textit{high} magnetic fields is nonetheless dominated by the presence of the Weyl nodes. Torque magnetometry up to 60 tesla reveals a change in the slope of $\tau/B$ at the quantum limit \Bs ($\approx 32\,\rm{T}$), where the chemical potential enters the $n=0$ Landau level. Numerical simulations show that this behaviour results from the magnetic field pulling the chemical potential to the chiral $n=0$ Landau level belonging to the Weyl nodes. These results show that high magnetic fields can uncover topological singularities in the underlying band structure of a potential WSM, and can recover topologically non-trivial experimental properties, even when the position of the chemical potential precludes their observation in zero magnetic field.}

\begin{figure}%
\includegraphics[width=\columnwidth]{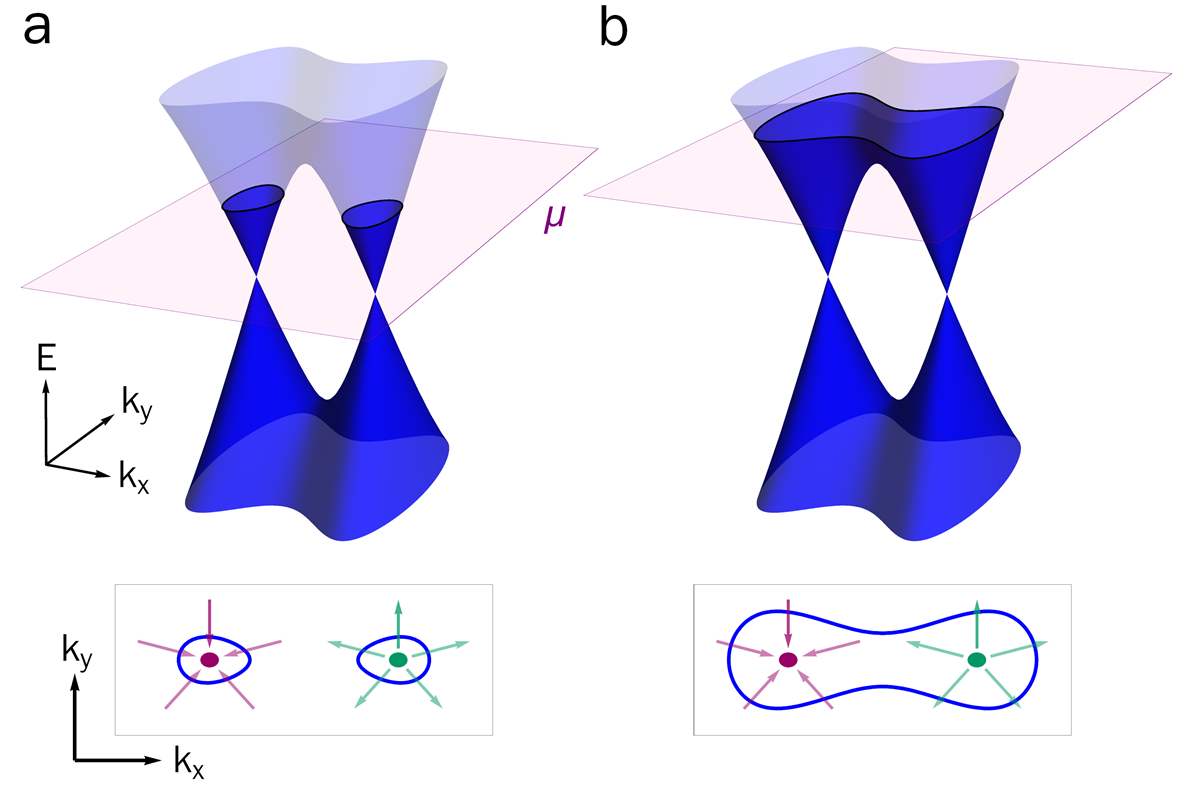}%
\caption{\textbf{The Fermi surface of a Weyl semimetal.} With the chemical potential sufficiently close to the nodes (\textbf{a}), the left and right-handed Weyl nodes form separate chiral Fermi surfaces (green and purple arrows indicate a source and sink of Berry flux). When the chemical potential lies above the saddle point the two chiral Fermi surfaces merge into a single surface with no net flux (\textbf{b}).}%
\label{fig:weyl}%
\end{figure}

Clear classification schemes have been put forward that unambiguously identify the topological character of a material from its electronic structure \cite{Thouless:1982,Ryu2010a}. Materials with bulk band gaps can be classified according to these schemes through the observation of topological surface states---topological insulators provide a prime example of this. The experimental situation in gapless topological systems is more complicated: depending on the position of the chemical potential, the topological band crossings may coexist with trivial quasiparticles or may lie too far from the chemical potential to be observed. This renders many potential Weyl or Dirac semimetals unsuitable for the study of bulk Weyl or Dirac Fermions. 

\autoref{fig:weyl} provides an intuitive example of how finely-tuned energy scales can determine the topological character of a WSM. When the chemical potential is sufficiently close to the Weyl nodes there exist separate Fermi surfaces of opposite chirality, ensuring the existence of Fermi-arc surface states \cite{Wan2011a} and of the chiral magnetic effect \cite{Nielsen,Son:2013,Hosur:2015}. A small shift in the chemical potential, however, merges the chiral Fermi surfaces and a larger, achiral Fermi surface appears. The Chern number of the resulting Fermi surface is zero, and thus the system will exhibit ``topologically trivial'' experimental properties despite the topology of its underlying bandstructure. These include a lack of the chiral magnetic effect, and zero Berry phase in quantum oscillation experiments.

Sensitivity to the fine-tuning of the chemical potential is prevalent in the inversion-symmetry broken Weyl-semimetals (Ta,Nb)(As,P) \cite{Weng2015,Huang2015}. As the spin-orbit interactions in these materials are small compared to overall electronic bandwidths, their Weyl nodes are not well-separated in momentum space and the energy barrier between them is only a few 10s of meV. Photoemission experiments find evidence for Weyl Fermi arcs in TaAs \cite{Xu2015a} and NbAs \cite{Xu2015}, indicating that the chemical potential is close to the nodes (\autoref{fig:weyl}a). In the isovalent and isostructural phosphides, however, the chemical potential appears to be positioned such that the Weyl nodes are encompassed by a single Fermi surface \cite{Kumar2017}. With zero net quasiparticle chirality, the phosphides have been argued to be topologically trivial \cite{Arnold:2015} (\autoref{fig:weyl}b).

\begin{figure}%
\includegraphics[width=1.95\columnwidth, trim=0.3cm 0cm 3.2cm 0.0cm, clip=true]{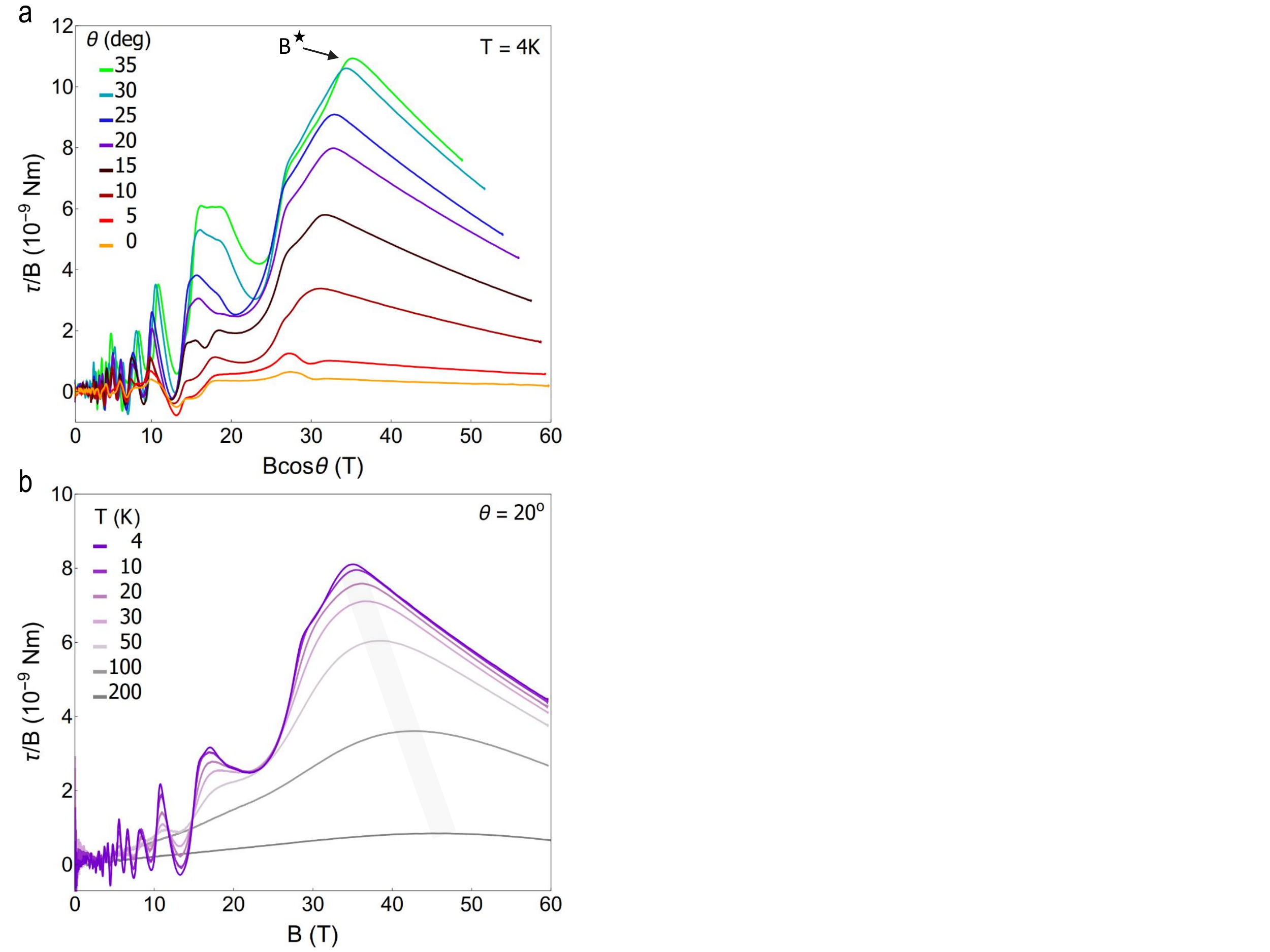}%
\caption{\textbf{Magnetic torque of NbP.} Panel \textbf{a} shows the angle dependence of $\tau/B$ at $T=4$~K. The angle $\theta$ is defined as the angle between the crystal $c$-axis and one of the tetragonal $a$-axes. We plot versus $B \cos \theta$ to show that the quantum oscillation frequencies scale roughly as $1/\cos\theta$ at low angle, consistent with the previously-determined ellipsoidal Fermi surface geometry of NbP \cite{Klotz:2016,Kumar2017}. \Bs, the field where the slope of $\tau/B$ changes sign, tracks the angle dependence of the quantum limit. (\textbf{b}) With field applied \degr{20} away from the $\hat{c}$-axis we measure the temperature dependence of $\tau/B$ from 4 to 200 K. \Bs increases with increasing temperature (grey band) and is still visible even at 200 K.}%
\label{fig:torque}%
\end{figure}

Applying a strong magnetic field introduces a new energy scale---the cyclotron energy---which modifies the electronic structure into Landau levels (LLs). Even states not located at the Fermi level, and thus invisible to most zero-field properties, contribute when the cyclotron energy is of order their distance from the chemical potential. Here we show how NbP, despite its achiral character in zero field, exhibits signatures of the underlying topological band structure in its magnetism in high magnetic fields. \autoref{fig:torque} shows the magnetic torque $\tau$ divided by magnetic field $\left(\vec{M}\!\times\!\vec{B}/|\vec{B}|\right)$ of NbP at multiple field orientations and temperatures. The torque follows the expected B$^2$ dependence at low fields, overlayed by strong de Haas-van Alphen oscillations. At base temperature this behavior continues up to the field \Bs, at which point a sharp kink and sudden decrease in torque is observed.

The Fermi surface of NbP consists of four sections: two nested sickle-shaped electron pockets, and two nested banana-shaped hole pockets \cite{Klotz:2016}. Torque measures the anisotropy of the Landau diamagnetism from each Fermi surface. Because each Fermi pocket is replicated symmetrically around the Brillouin zone such that time reversal symmetry is preserved, torque disappears with the field applied along a crystallographic direction. With the field applied near the $\hat{c}$-axis the quasiparticles form ``belly'' cyclotron orbits around both the electron and hole Fermi surfaces. The last quantum oscillation occurs at 32 tesla \cite{Shekhar2015}---this is known as the quantum limit, where the chemical potential enters the $n=0$ LL. The quantum limit coincides with the kink-field \Bs, and this field scale increases as the magnetic field is titled away from the $c$-axis (\autoref{fig:torque}a).

The energy of the $n=0$ LL in a trivial parabolic band increases linearly with magnetic field, dispersing as
\begin{equation}
E_0 = \frac{1}{2} \hbar \frac{e B}{m^{\star}_{\perp}} + \frac{\hbar^2 k_z^2}{2 m_z^{\star}},
\label{eq:triv}
\end{equation}
where $m^{\star}_{\perp}$ is the orbitally-averaged cyclotron mass perpendicular to the magnetic field, and $m_z^{\star}$ and $k_z$ are the mass and momentum along the field direction. In Weyl systems, on the other hand, each node has an $n=0$ LL that is field independent and disperses linearly in $k_z$:
\begin{equation}
E_0 = \pm \hbar v_z k_z,
\label{eq:weyl}
\end{equation}
where $v_z$ is the Fermi velocity along the field direction and the positive (negative) sign denotes the right (left) Weyl node. It was previously shown that a change in slope of $\tau/B$ at the quantum limit is associated with the chemical potential moving to the field-independent $n=0$ LL \cite{Moll2016}. This is because, while the $n>0$ LLs in both trivial and Weyl systems are field-dependent, the  $n=0$ LL of Weyl systems is field-independent. As magnetization (proportional to $\tau/B$) is the derivative of the free energy with respect to magnetic field, there is a qualitative difference between trivial and Weyl systems at the quantum limit. Here, a sharp change in torque is associated with the loss of competition between field-dependent $n>0$ and field-independent $n=0$ LLs. This phenomenon has been observed in the closely related Weyl semimetal NbAs, where separate Fermi surfaces exist for left- and right-handed Weyl Fermions in zero magnetic field \cite{Moll2016}. 

\begin{figure}[H]%
\includegraphics[width=\columnwidth, trim=0cm 0cm 14cm 0.0cm, clip=true]{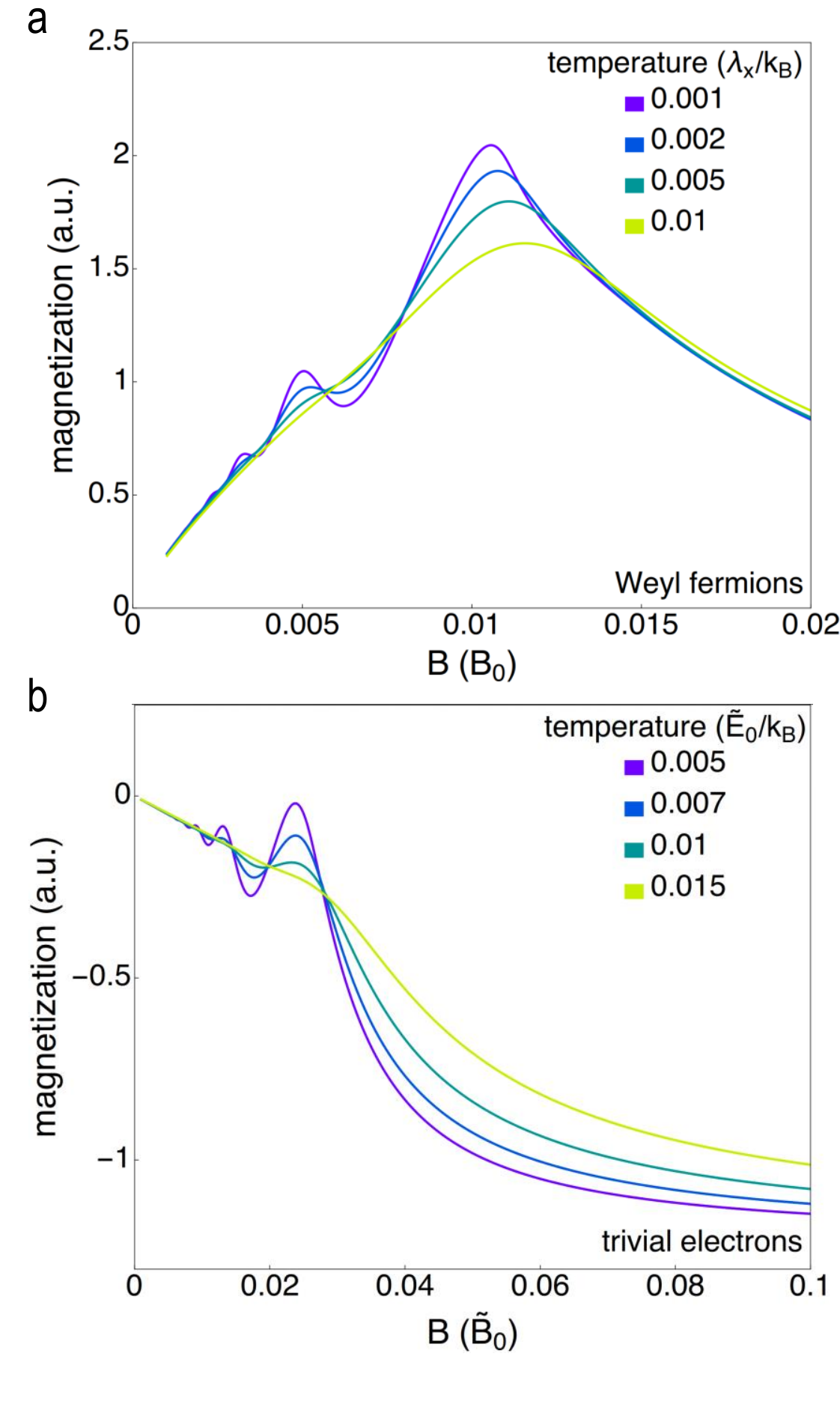}
\caption{\textbf{Simulated magnetization of Weyl Fermions and trivial quasiparticles.} Panel \textbf{a} depicts the magnetization as a function of field for different temperatures. While the magnetization (which is proportional to $\tau/B$) increases at small fields due to the diamagnetic contribution of the $n>0$ Landau levels, it decreases when the chemical potential enters the $n=0$ Landau level. As the temperature is increased the magnetization is washed out, and the field at which the last peak occurs, $\Bs$,  shifts to larger values. Panel \textbf{b} contrasts this with the magnetization of trivial Landau levels, which approaches a finite and constant value at large fields (here, magnetic fields are measured in units of $\tilde{B}_0= 2\tilde{E}_0 m_\perp/\hbar e$ and the chemical potential at zero field is $\mu = 0.06\,\tilde{E}_0$, where $E_0$ are units of energy). 
}%
\label{fig:magnetization}%
\end{figure}

In NbP, the chemical potential lies above the saddle point separating the Weyl nodes at zero magnetic field (\autoref{fig:weyl}b). The fact that we observe a change in slope in $\tau/B$ at \Bs signifies that the magnetic field shifts the chemical potential to the chiral $n=0$ Landau level above the quantum limit. We find that \Bs is still visible at 200 K, but has shifted to higher magnetic field (\autoref{fig:torque}b). We understand the increase in \Bs with temperature as the thermal population of the diamagnetic $n=1$ Landau level, which competes with the $n=0$ Landau level. The temperature dependence of \Bs is also confirmed in our numerical calculations (Figure \ref{fig:magnetization}).

We model our data using a time-reversal symmetric Weyl band structure whose chemical potential is above the saddle point in zero magnetic field. To analyze the dependence of the magnetization ($\propto \tau/B$) on magnetic field strength and temperature, we employ a minimal tight-binding model with two atoms per unit cell \cite{chen:2016}. Denoting the physical spin by $\mathbf{\sigma}$ and the sublattice pseudospin by $\mathbf{\tau}$, the zero-field Hamiltonian is given by $H=\sum_{\mathbf{k}}\Psi_\mathbf{k}^\dagger\mathcal{H}_\mathbf{k}\Psi_\mathbf{k}^\pd$ with
\begin{align}
%H&=\sum_{\mathbf{k}}\Psi_\mathbf{k}^\dagger\mathcal{H}_\mathbf{k}\Psi_\mathbf{k}^\pd,\\
\mathcal{H}_\mathbf{k}&=\lambda_x\sin(k_xa)\,\sigma_x\mathds{1}_\tau+\sum_{j=y,z}\lambda_{yz}\sin(k_ja)\,\sigma_j\mathds{1}_\tau\nonumber\\
&+M(\mathbf{k})\,\sigma_x\tau_x-\mu\,\mathds{1}_\sigma\mathds{1}_\tau,
\end{align}
where $\Psi_\mathbf{k}^\dagger =(c_{\mathbf{k},\uparrow,+}^\dagger,c_{\mathbf{k},\downarrow,+}^\dagger,c_{\mathbf{k},\uparrow,-}^\dagger,c_{\mathbf{k},\downarrow,-}^\dagger)$ contains the creation operators for electrons of momentum $\hbar\mathbf{k}$ and spin $\sigma=\uparrow,\downarrow$ on sublattice $\tau=\pm$, and $M(\mathbf{k})=m+m'(2-\cos(k_ya)-\cos(k_za))$ controls the separation of the Weyl nodes. In these expressions, $\lambda_x$, $\lambda_{yz}$, $m$, and $m'$ denote different hopping energies, $a$ is the lattice constant, and $\mu$ is the chemical potential. This model has time-reversal symmetry $\mathcal{T}=\mathcal{K}i\sigma_y\tau_z$ ($\mathcal{K}$ denotes complex conjugation). In the remainder, we focus on the regime $2m'+m>\lambda_x>m>0$, in which the low-energy physics is due to four Weyl nodes on the $k_x$-axis. For $\lambda_{yz}\gg\lambda_x$, they are described by the expansion of the Hamiltonian to linear order in $k_{y}$ and $k_{z}$,
\begin{align}
\mathcal{H}_\mathbf{k}&\approx\lambda_x\,\sin(k_xa)\,\sigma_x\mathds{1}_\tau+m\,\sigma_x\,\tau_x\nonumber\\
&+\sum_{j=y,z}\lambda_{yz}\,a\,k_j\,\sigma_j\mathds{1}_\tau-\mu\,\mathds{1}_\sigma\mathds{1}_\tau
\end{align}

\begin{figure}%
\includegraphics[scale=0.35]{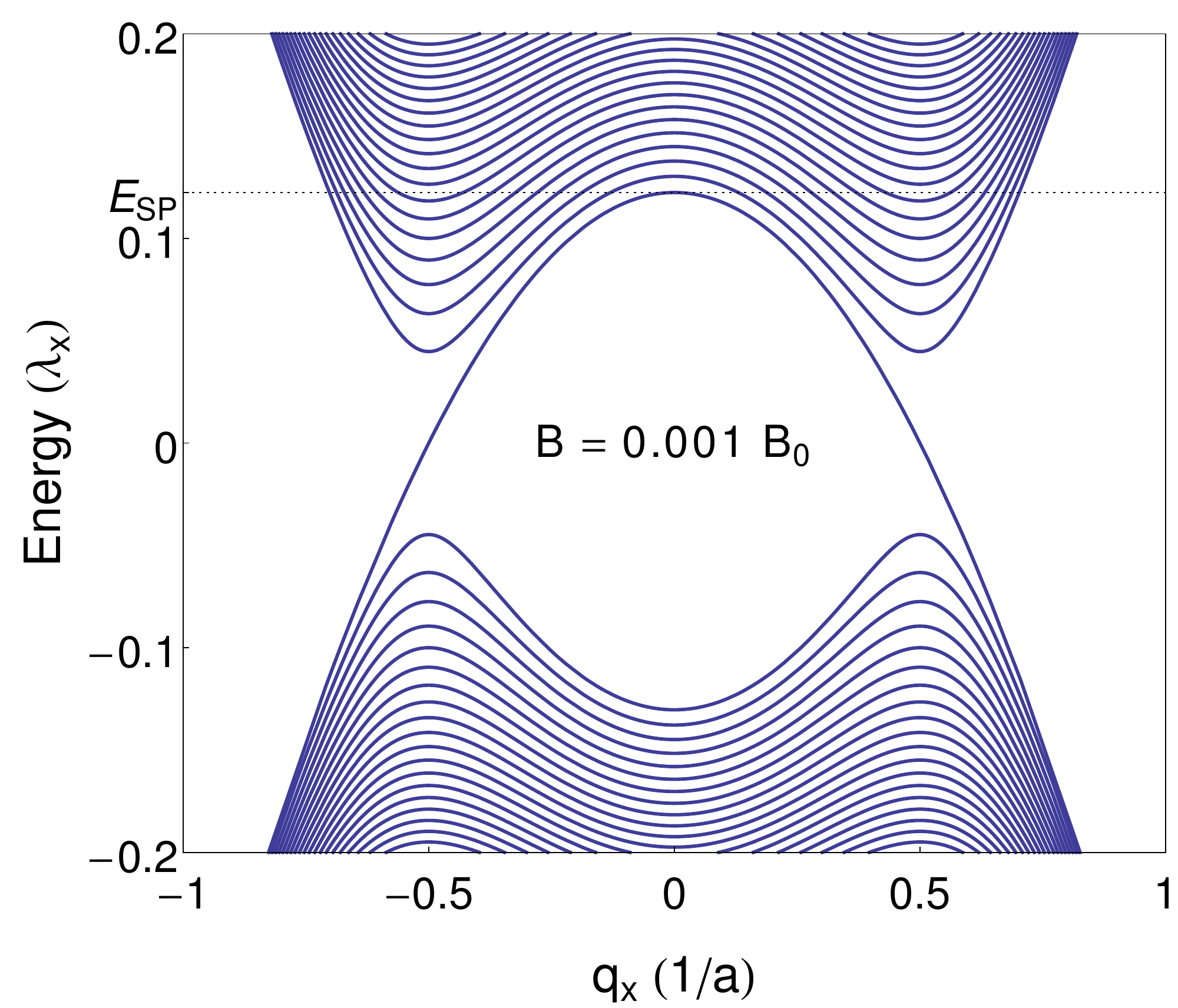}%
\caption{\textbf{Landau levels for one sector.} Landau level spectrum of one of the two time-reversal partner sectors in our tight-binding model as a function of the (shifted) wave number $q_x=k_x+\pi/2$ for a magnetic field of amplitude $B=0.001\,B_0$ along the $x$-direction (where $B_0 = \hbar \lambda_x^2/\lambda_{yz}^2 a^2e$). At zero field, the model features Weyl nodes at $q_x=\pm 0.5$ and zero energy. The dotted horizontal line indicates the saddle point energy $E_{\rm SP}$. The spectrum of the time-reversal partner is particle-hole symmetric to this one.}%
\label{fig:spectrum}%
\end{figure}

We now calculate the Landau levels for a magnetic field applied along the $x$-axis (the field strength is measured in units of $B_0 = \hbar \lambda_x^2/\lambda_{yz}^2 a^2e$). The total free energy is then computed at fixed quasiparticle density by adjusting the chemical potential accordingly as a function of magnetic field. This corresponds to the physical situation in a three dimensional metal, where charge neutrality constrains the number of carriers and the chemical potential oscillates with magnetic field. The quasiparticle number is given as the number of occupied single-particle states above $E=0$ in both time-reversal partner sectors combined, plus a constant corresponding to the filled valence band (this constant is independent of the magnetic field because the two time-reversal partner sectors have particle-hole symmetric spectra).

\begin{figure}%
\includegraphics[width=2\columnwidth, trim=0.3cm 0cm 2.0cm 0.0cm, clip=true]{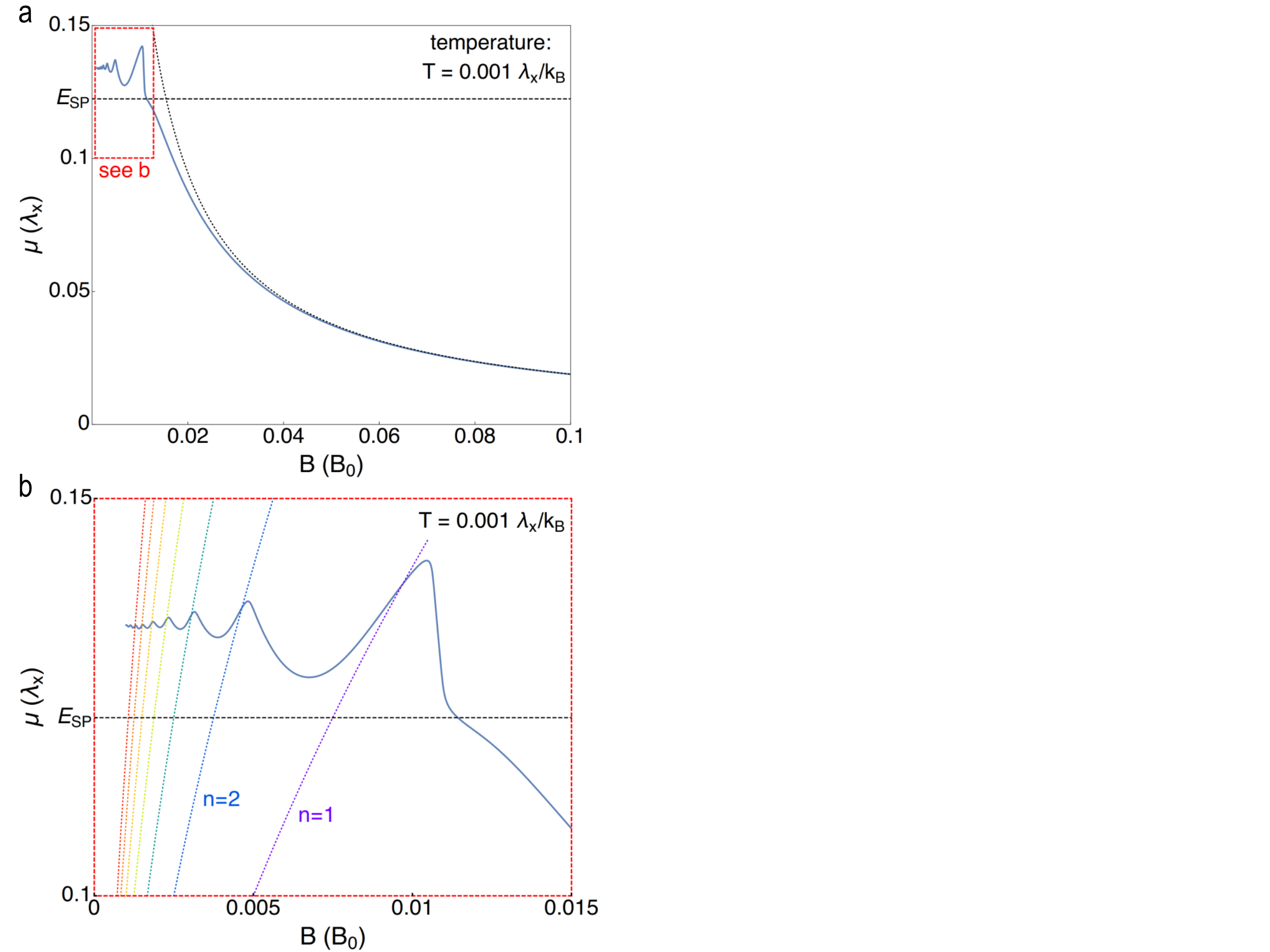}%
\caption{\textbf{Magnetic field dependence of the chemical potential in a Weyl semimetal.} Panel \textbf{a} shows field dependence of the chemical potential. At zero field, the Weyl nodes are located at zero energy, and the horizontal dashed line indicates $E_{\rm SP}$, the saddle point energy above which the Fermi surface encloses an equal number of nodes of each chirality (the topologically trivial regime). At large fields, the chemical potential is located within the $n=0$ Landau level, and approaches $\mu(B) = \pi^2 \rho v_F/B$ (where $\rho$ is the quasiparticle density, and $v_F$ denotes the Fermi velocity, this is shown by the dotted line). Panel \textbf{b} is a zoom into the low-field region. The curved dotted lines depict the energies of the bottoms of the higher Landau levels with $n=1$ through $n=7$.}%
\label{fig:chem_pot}%
\end{figure}

\autoref{fig:chem_pot} shows that the chemical potential is roughly constant at small fields, with quantum oscillations appearing as the chemical potential adjusts to keep carrier number fixed. Beyond $\Bs$, the field strength at which the $n=1$ Landau level is depopulated at zero temperature, the large degeneracy of the $n=0$ LL allows it to accommodate all quasiparticles. While the energy of the $n=0$ LL is field-independent, its degeneracy increases linearly in $B$, pulling the chemical potential toward the Weyl nodes at $E=0$ as $B \rightarrow \infty$ (\autoref{fig:chem_pot}a). 

We compute the magnetization as the field derivative of the free energy, $M = - \frac{dF}{dB}$. The $n>0$ LLs increase in energy with increasing field, contributing a diamagnetic response to the magnetization. The $n=0$ LL, on the flip side, contributes a paramagnetic response as the chemical potential moves towards $E=0$ in the quantum limit. As a result, the slope of the magnetization changes sign when the system enters the quantum limit, see Fig.~\ref{fig:magnetization}{a}. When the temperature increases, we observe a shift of the final maximum toward larger fields due to the thermal population of the $n=1$ Landau level. Both the field dependence and the temperature dependence of the computed magnetization are in good agreement with the experimental data.

This behavior can be contrasted to the magnetization of a free electron gas with parabolic dispersion. The large degeneracy of the LLs again implies that when the total carrier density is held constant, all quasiparticles bunch at the bottom of the $n=0$ LL at large fields. Because the energies of all parabolic LLs (including the $n=0$ LL) increase with field, however, the energy density is now asymptotically given by $\epsilon = \rho \hbar \omega_c/2$, where $\omega_c = \frac{eB}{m^{\star}}$ is the cyclotron frequency and $\rho$ is the quasiparticle density. In stark contrast to the experimentally observed behavior in NbP, the large-field magnetization for a trivial band approaches a constant finite value.

Weyl Fermions exhibit unusual transport and optical properties due to the chiral anomaly \cite{Nielsen,Son:2013,Hosur:2015,Lucas:2016,Ramshaw:2017} and the mixed axial-gravitational anomaly \cite{Lucas:2016,Gooth:2017}, and are the starting point for several predicted chiral states of matter \cite{Wei:2012,Cho:2012,Meng:2012,Zhang:2017b}. Metals hosting Weyl nodes ought to be relatively common: in three dimensions the band crossings that give rise to linear energy-momentum dispersions do not require the fine tuning that they require in two dimensions \cite{herring:1937,Vafek:2014}. Like NbP, several candidate materials have the chemical potential crossing far from the Weyl nodes forming topologically trivial Fermi surfaces in zero magnetic field. We have shown that high magnetic fields can be used to shift the chemical potential to the fully chiral $n=0$ Landau level in the quantum limit, greatly broadening the number of systems where Weyl Fermions can be studied in practice. The same high magnetic fields also increase the Coulomb interaction between quasiparticles (generally weak in semimetals due to their high Fermi velocities), driving these systems closer to new symmetry-breaking or interacting topological states of matter.

\section*{Acknowledgments}
The authors wish to thank A. Shekhter useful discussions. B.J.R.~acknowledges funding from LANL LDRD 20160616ECR `New States of Matter in Weyl Semimetals'. T.M.~is funded by Deutsche Forschungsgemeinschaft through GRK 1621, SFB 1143, and the Emmy-Noether program ME 4844/1. P.M. is supported by the Max-Planck-Society and the European Research Council (ERC) under the European Union's Horizon 2020 research and innovation programme (grant agreement No. 715730).

\section*{Author Contributions}
B.J.R., K.A.M., and P.J.W.M. conceived of the experiment; E.D.B. and F.R. grew the samples, B.J.R. and K.A.M. performed the experiment and analyzed the data; T.M. performed the numerical simulations; B.J.R. and K.A.M. wrote the manuscript with contributions from all coauthors.

%Additionally, the $n>0$ Landau levels at low field do not have well-defined chirality even with the chemical potential crossing close to the nodes \cite{Arnold:2016b}. 

%\bibliographystyle{unsrtnat}
\bibliography{NbPrefs}

\begin{thebibliography}{25}%
\makeatletter
\providecommand \@ifxundefined [1]{%
 \@ifx{#1\undefined}
}%
\providecommand \@ifnum [1]{%
 \ifnum #1\expandafter \@firstoftwo
 \else \expandafter \@secondoftwo
 \fi
}%
\providecommand \@ifx [1]{%
 \ifx #1\expandafter \@firstoftwo
 \else \expandafter \@secondoftwo
 \fi
}%
\providecommand \natexlab [1]{#1}%
\providecommand \enquote  [1]{``#1''}%
\providecommand \bibnamefont  [1]{#1}%
\providecommand \bibfnamefont [1]{#1}%
\providecommand \citenamefont [1]{#1}%
\providecommand \href@noop [0]{\@secondoftwo}%
\providecommand \href [0]{\begingroup \@sanitize@url \@href}%
\providecommand \@href[1]{\@@startlink{#1}\@@href}%
\providecommand \@@href[1]{\endgroup#1\@@endlink}%
\providecommand \@sanitize@url [0]{\catcode `\\12\catcode `\$12\catcode
  `\&12\catcode `\#12\catcode `\^12\catcode `\_12\catcode `\%12\relax}%
\providecommand \@@startlink[1]{}%
\providecommand \@@endlink[0]{}%
\providecommand \url  [0]{\begingroup\@sanitize@url \@url }%
\providecommand \@url [1]{\endgroup\@href {#1}{\urlprefix }}%
\providecommand \urlprefix  [0]{URL }%
\providecommand \Eprint [0]{\href }%
\providecommand \doibase [0]{http://dx.doi.org/}%
\providecommand \selectlanguage [0]{\@gobble}%
\providecommand \bibinfo  [0]{\@secondoftwo}%
\providecommand \bibfield  [0]{\@secondoftwo}%
\providecommand \translation [1]{[#1]}%
\providecommand \BibitemOpen [0]{}%
\providecommand \bibitemStop [0]{}%
\providecommand \bibitemNoStop [0]{.\EOS\space}%
\providecommand \EOS [0]{\spacefactor3000\relax}%
\providecommand \BibitemShut  [1]{\csname bibitem#1\endcsname}%
\let\auto@bib@innerbib\@empty
%</preamble>
\bibitem [{\citenamefont {Thouless}\ \emph {et~al.}(1982)\citenamefont
  {Thouless}, \citenamefont {Kohmoto}, \citenamefont {Nightingale},\ and\
  \citenamefont {den Nijs}}]{Thouless:1982}%
  \BibitemOpen
  \bibfield  {author} {\bibinfo {author} {\bibfnamefont {D.~J.}\ \bibnamefont
  {Thouless}}, \bibinfo {author} {\bibfnamefont {M.}~\bibnamefont {Kohmoto}},
  \bibinfo {author} {\bibfnamefont {M.~P.}\ \bibnamefont {Nightingale}}, \ and\
  \bibinfo {author} {\bibfnamefont {M.}~\bibnamefont {den Nijs}},\ }\href
  {\doibase 10.1103/PhysRevLett.49.405} {\bibfield  {journal} {\bibinfo
  {journal} {Phys. Rev. Lett.}\ }\textbf {\bibinfo {volume} {49}},\ \bibinfo
  {pages} {405} (\bibinfo {year} {1982})}\BibitemShut {NoStop}%
\bibitem [{\citenamefont {Ryu}\ \emph {et~al.}(2010)\citenamefont {Ryu},
  \citenamefont {Schnyder}, \citenamefont {Furusaki},\ and\ \citenamefont
  {Ludwig}}]{Ryu2010a}%
  \BibitemOpen
  \bibfield  {author} {\bibinfo {author} {\bibfnamefont {S.}~\bibnamefont
  {Ryu}}, \bibinfo {author} {\bibfnamefont {A.~P.}\ \bibnamefont {Schnyder}},
  \bibinfo {author} {\bibfnamefont {A.}~\bibnamefont {Furusaki}}, \ and\
  \bibinfo {author} {\bibfnamefont {A.~W.~W.}\ \bibnamefont {Ludwig}},\ }\href
  {\doibase 10.1088/1367-2630/12/6/065010} {\bibfield  {journal} {\bibinfo
  {journal} {New Journal of Physics}\ }\textbf {\bibinfo {volume} {12}}
  (\bibinfo {year} {2010}),\ 10.1088/1367-2630/12/6/065010},\ \Eprint
  {http://arxiv.org/abs/0912.2157} {arXiv:0912.2157} \BibitemShut {NoStop}%
\bibitem [{\citenamefont {Wan}\ \emph {et~al.}(2011)\citenamefont {Wan},
  \citenamefont {Turner}, \citenamefont {Vishwanath},\ and\ \citenamefont
  {Savrasov}}]{Wan2011a}%
  \BibitemOpen
  \bibfield  {author} {\bibinfo {author} {\bibfnamefont {X.}~\bibnamefont
  {Wan}}, \bibinfo {author} {\bibfnamefont {A.~M.}\ \bibnamefont {Turner}},
  \bibinfo {author} {\bibfnamefont {A.}~\bibnamefont {Vishwanath}}, \ and\
  \bibinfo {author} {\bibfnamefont {S.~Y.}\ \bibnamefont {Savrasov}},\ }\href
  {\doibase 10.1103/PhysRevB.83.205101} {\bibfield  {journal} {\bibinfo
  {journal} {Physical Review B - Condensed Matter and Materials Physics}\
  }\textbf {\bibinfo {volume} {83}} (\bibinfo {year} {2011}),\
  10.1103/PhysRevB.83.205101},\ \Eprint {http://arxiv.org/abs/1007.0016}
  {arXiv:1007.0016} \BibitemShut {NoStop}%
\bibitem [{\citenamefont {Nielsen}\ and\ \citenamefont
  {Ninomiya}(1983)}]{Nielsen}%
  \BibitemOpen
  \bibfield  {author} {\bibinfo {author} {\bibfnamefont {H.~B.}\ \bibnamefont
  {Nielsen}}\ and\ \bibinfo {author} {\bibfnamefont {M.}~\bibnamefont
  {Ninomiya}},\ }\href {\doibase 10.1016/0370-2693(83)91529-0} {\bibfield
  {journal} {\bibinfo  {journal} {Phys. Lett. B}\ } (\bibinfo {year} {1983}),\
  10.1016/0370-2693(83)91529-0}\BibitemShut {NoStop}%
\bibitem [{\citenamefont {Son}\ and\ \citenamefont {Spivak}(2013)}]{Son:2013}%
  \BibitemOpen
  \bibfield  {author} {\bibinfo {author} {\bibfnamefont {D.~T.}\ \bibnamefont
  {Son}}\ and\ \bibinfo {author} {\bibfnamefont {B.~Z.}\ \bibnamefont
  {Spivak}},\ }\href {\doibase 10.1103/PhysRevB.88.104412} {\bibfield
  {journal} {\bibinfo  {journal} {Phys. Rev. B}\ }\textbf {\bibinfo {volume}
  {88}},\ \bibinfo {pages} {104412} (\bibinfo {year} {2013})}\BibitemShut
  {NoStop}%
\bibitem [{\citenamefont {Hosur}\ and\ \citenamefont {Qi}(2015)}]{Hosur:2015}%
  \BibitemOpen
  \bibfield  {author} {\bibinfo {author} {\bibfnamefont {P.}~\bibnamefont
  {Hosur}}\ and\ \bibinfo {author} {\bibfnamefont {X.-L.}\ \bibnamefont {Qi}},\
  }\href@noop {} {\bibfield  {journal} {\bibinfo  {journal} {Physical Review
  B}\ }\textbf {\bibinfo {volume} {91}},\ \bibinfo {pages} {081106} (\bibinfo
  {year} {2015})}\BibitemShut {NoStop}%
\bibitem [{\citenamefont {Weng}\ \emph {et~al.}(2015)\citenamefont {Weng},
  \citenamefont {Fang}, \citenamefont {Fang}, \citenamefont {Bernevig},\ and\
  \citenamefont {Dai}}]{Weng2015}%
  \BibitemOpen
  \bibfield  {author} {\bibinfo {author} {\bibfnamefont {H.}~\bibnamefont
  {Weng}}, \bibinfo {author} {\bibfnamefont {C.}~\bibnamefont {Fang}}, \bibinfo
  {author} {\bibfnamefont {Z.}~\bibnamefont {Fang}}, \bibinfo {author}
  {\bibfnamefont {B.~A.}\ \bibnamefont {Bernevig}}, \ and\ \bibinfo {author}
  {\bibfnamefont {X.}~\bibnamefont {Dai}},\ }\href {\doibase
  10.1103/PhysRevX.5.011029} {\bibfield  {journal} {\bibinfo  {journal} {Phys.
  Rev. X}\ }\textbf {\bibinfo {volume} {5}},\ \bibinfo {pages} {011029}
  (\bibinfo {year} {2015})}\BibitemShut {NoStop}%
\bibitem [{\citenamefont {Huang}\ \emph {et~al.}(2015)\citenamefont {Huang},
  \citenamefont {Xu}, \citenamefont {Belopolski}, \citenamefont {Lee},
  \citenamefont {Chang}, \citenamefont {Wang}, \citenamefont {Alidoust},
  \citenamefont {Bian}, \citenamefont {Neupane}, \citenamefont {Zhang},
  \citenamefont {Jia}, \citenamefont {Bansil}, \citenamefont {Lin},\ and\
  \citenamefont {Hasan}}]{Huang2015}%
  \BibitemOpen
  \bibfield  {author} {\bibinfo {author} {\bibfnamefont {S.-M.}\ \bibnamefont
  {Huang}}, \bibinfo {author} {\bibfnamefont {S.-Y.}\ \bibnamefont {Xu}},
  \bibinfo {author} {\bibfnamefont {I.}~\bibnamefont {Belopolski}}, \bibinfo
  {author} {\bibfnamefont {C.-C.}\ \bibnamefont {Lee}}, \bibinfo {author}
  {\bibfnamefont {G.}~\bibnamefont {Chang}}, \bibinfo {author} {\bibfnamefont
  {B.}~\bibnamefont {Wang}}, \bibinfo {author} {\bibfnamefont {N.}~\bibnamefont
  {Alidoust}}, \bibinfo {author} {\bibfnamefont {G.}~\bibnamefont {Bian}},
  \bibinfo {author} {\bibfnamefont {M.}~\bibnamefont {Neupane}}, \bibinfo
  {author} {\bibfnamefont {C.}~\bibnamefont {Zhang}}, \bibinfo {author}
  {\bibfnamefont {S.}~\bibnamefont {Jia}}, \bibinfo {author} {\bibfnamefont
  {A.}~\bibnamefont {Bansil}}, \bibinfo {author} {\bibfnamefont
  {H.}~\bibnamefont {Lin}}, \ and\ \bibinfo {author} {\bibfnamefont {M.~Z.}\
  \bibnamefont {Hasan}},\ }\href {\doibase 10.1038/ncomms8373} {\bibfield
  {journal} {\bibinfo  {journal} {Nature Communications}\ }\textbf {\bibinfo
  {volume} {6}},\ \bibinfo {pages} {7373} (\bibinfo {year} {2015})}\BibitemShut
  {NoStop}%
\bibitem [{\citenamefont {Xu}\ \emph {et~al.}(2015{\natexlab{a}})\citenamefont
  {Xu}, \citenamefont {Belopolski}, \citenamefont {Alidoust}, \citenamefont
  {Neupane}, \citenamefont {Bian}, \citenamefont {Zhang}, \citenamefont
  {Sankar}, \citenamefont {Chang}, \citenamefont {Yuan}, \citenamefont {Lee},
  \citenamefont {Huang}, \citenamefont {Zheng}, \citenamefont {Ma},
  \citenamefont {Sanchez}, \citenamefont {Wang}, \citenamefont {Bansil},
  \citenamefont {Chou}, \citenamefont {Shibayev}, \citenamefont {Lin},
  \citenamefont {Jia},\ and\ \citenamefont {Hasan}}]{Xu2015a}%
  \BibitemOpen
  \bibfield  {author} {\bibinfo {author} {\bibfnamefont {S.-Y.}\ \bibnamefont
  {Xu}}, \bibinfo {author} {\bibfnamefont {I.}~\bibnamefont {Belopolski}},
  \bibinfo {author} {\bibfnamefont {N.}~\bibnamefont {Alidoust}}, \bibinfo
  {author} {\bibfnamefont {M.}~\bibnamefont {Neupane}}, \bibinfo {author}
  {\bibfnamefont {G.}~\bibnamefont {Bian}}, \bibinfo {author} {\bibfnamefont
  {C.}~\bibnamefont {Zhang}}, \bibinfo {author} {\bibfnamefont
  {R.}~\bibnamefont {Sankar}}, \bibinfo {author} {\bibfnamefont
  {G.}~\bibnamefont {Chang}}, \bibinfo {author} {\bibfnamefont
  {Z.}~\bibnamefont {Yuan}}, \bibinfo {author} {\bibfnamefont {C.-C.}\
  \bibnamefont {Lee}}, \bibinfo {author} {\bibfnamefont {S.-M.}\ \bibnamefont
  {Huang}}, \bibinfo {author} {\bibfnamefont {H.}~\bibnamefont {Zheng}},
  \bibinfo {author} {\bibfnamefont {J.}~\bibnamefont {Ma}}, \bibinfo {author}
  {\bibfnamefont {D.~S.}\ \bibnamefont {Sanchez}}, \bibinfo {author}
  {\bibfnamefont {B.}~\bibnamefont {Wang}}, \bibinfo {author} {\bibfnamefont
  {A.}~\bibnamefont {Bansil}}, \bibinfo {author} {\bibfnamefont
  {F.}~\bibnamefont {Chou}}, \bibinfo {author} {\bibfnamefont {P.~P.}\
  \bibnamefont {Shibayev}}, \bibinfo {author} {\bibfnamefont {H.}~\bibnamefont
  {Lin}}, \bibinfo {author} {\bibfnamefont {S.}~\bibnamefont {Jia}}, \ and\
  \bibinfo {author} {\bibfnamefont {M.~Z.}\ \bibnamefont {Hasan}},\ }\href
  {\doibase 10.1126/science.aaa9297} {\bibfield  {journal} {\bibinfo  {journal}
  {Science}\ }\textbf {\bibinfo {volume} {349}},\ \bibinfo {pages} {613}
  (\bibinfo {year} {2015}{\natexlab{a}})},\ \Eprint
  {http://arxiv.org/abs/1502.03807} {arXiv:1502.03807} \BibitemShut {NoStop}%
\bibitem [{\citenamefont {Xu}\ \emph {et~al.}(2015{\natexlab{b}})\citenamefont
  {Xu}, \citenamefont {Alidoust}, \citenamefont {Belopolski}, \citenamefont
  {Yuan}, \citenamefont {Bian}, \citenamefont {Chang}, \citenamefont {Zheng},
  \citenamefont {Strocov}, \citenamefont {Sanchez}, \citenamefont {Chang},
  \citenamefont {Zhang}, \citenamefont {Mou}, \citenamefont {Wu}, \citenamefont
  {Huang}, \citenamefont {Lee},\ and\ \citenamefont {Huang}}]{Xu2015}%
  \BibitemOpen
  \bibfield  {author} {\bibinfo {author} {\bibfnamefont {S.-y.}\ \bibnamefont
  {Xu}}, \bibinfo {author} {\bibfnamefont {N.}~\bibnamefont {Alidoust}},
  \bibinfo {author} {\bibfnamefont {I.}~\bibnamefont {Belopolski}}, \bibinfo
  {author} {\bibfnamefont {Z.}~\bibnamefont {Yuan}}, \bibinfo {author}
  {\bibfnamefont {G.}~\bibnamefont {Bian}}, \bibinfo {author} {\bibfnamefont
  {T.-r.}\ \bibnamefont {Chang}}, \bibinfo {author} {\bibfnamefont
  {H.}~\bibnamefont {Zheng}}, \bibinfo {author} {\bibfnamefont {V.~N.}\
  \bibnamefont {Strocov}}, \bibinfo {author} {\bibfnamefont {D.~S.}\
  \bibnamefont {Sanchez}}, \bibinfo {author} {\bibfnamefont {G.}~\bibnamefont
  {Chang}}, \bibinfo {author} {\bibfnamefont {C.}~\bibnamefont {Zhang}},
  \bibinfo {author} {\bibfnamefont {D.}~\bibnamefont {Mou}}, \bibinfo {author}
  {\bibfnamefont {Y.}~\bibnamefont {Wu}}, \bibinfo {author} {\bibfnamefont
  {L.}~\bibnamefont {Huang}}, \bibinfo {author} {\bibfnamefont {C.-c.}\
  \bibnamefont {Lee}}, \ and\ \bibinfo {author} {\bibfnamefont {S.-m.}\
  \bibnamefont {Huang}},\ }\href {\doibase 10.1038/NPHYS3437} {\bibfield
  {journal} {\bibinfo  {journal} {Nature Physics}\ }\textbf {\bibinfo {volume}
  {11}},\ \bibinfo {pages} {748} (\bibinfo {year}
  {2015}{\natexlab{b}})}\BibitemShut {NoStop}%
\bibitem [{\citenamefont {Kumar}\ \emph {et~al.}(2017)\citenamefont {Kumar},
  \citenamefont {Neha}, \citenamefont {Das},\ and\ \citenamefont
  {Patnaik}}]{Kumar2017}%
  \BibitemOpen
  \bibfield  {author} {\bibinfo {author} {\bibfnamefont {P.}~\bibnamefont
  {Kumar}}, \bibinfo {author} {\bibfnamefont {P.}~\bibnamefont {Neha}},
  \bibinfo {author} {\bibfnamefont {T.}~\bibnamefont {Das}}, \ and\ \bibinfo
  {author} {\bibfnamefont {S.}~\bibnamefont {Patnaik}},\ }\href {\doibase
  10.1038/srep46062} {\bibfield  {journal} {\bibinfo  {journal} {Sci. Rep.}\ }
  (\bibinfo {year} {2017}),\ 10.1038/srep46062}\BibitemShut {NoStop}%
\bibitem [{\citenamefont {Arnold}\ \emph {et~al.}(2016)\citenamefont {Arnold},
  \citenamefont {Shekhar}, \citenamefont {Wu}, \citenamefont {Sun},
  \citenamefont {dos Reis}, \citenamefont {Kumar}, \citenamefont {Naumann},
  \citenamefont {Ajeesh}, \citenamefont {Schmidt}, \citenamefont {Grushin},
  \citenamefont {Bardarson}, \citenamefont {Baenitz}, \citenamefont {Sokolov},
  \citenamefont {Borrmann}, \citenamefont {Nicklas}, \citenamefont {Felser},
  \citenamefont {Hassinger},\ and\ \citenamefont {Yan}}]{Arnold:2015}%
  \BibitemOpen
  \bibfield  {author} {\bibinfo {author} {\bibfnamefont {F.}~\bibnamefont
  {Arnold}}, \bibinfo {author} {\bibfnamefont {C.}~\bibnamefont {Shekhar}},
  \bibinfo {author} {\bibfnamefont {S.-C.}\ \bibnamefont {Wu}}, \bibinfo
  {author} {\bibfnamefont {Y.}~\bibnamefont {Sun}}, \bibinfo {author}
  {\bibfnamefont {R.~D.}\ \bibnamefont {dos Reis}}, \bibinfo {author}
  {\bibfnamefont {N.}~\bibnamefont {Kumar}}, \bibinfo {author} {\bibfnamefont
  {M.}~\bibnamefont {Naumann}}, \bibinfo {author} {\bibfnamefont {M.~O.}\
  \bibnamefont {Ajeesh}}, \bibinfo {author} {\bibfnamefont {M.}~\bibnamefont
  {Schmidt}}, \bibinfo {author} {\bibfnamefont {A.~G.}\ \bibnamefont
  {Grushin}}, \bibinfo {author} {\bibfnamefont {J.~H.}\ \bibnamefont
  {Bardarson}}, \bibinfo {author} {\bibfnamefont {M.}~\bibnamefont {Baenitz}},
  \bibinfo {author} {\bibfnamefont {D.}~\bibnamefont {Sokolov}}, \bibinfo
  {author} {\bibfnamefont {H.}~\bibnamefont {Borrmann}}, \bibinfo {author}
  {\bibfnamefont {M.}~\bibnamefont {Nicklas}}, \bibinfo {author} {\bibfnamefont
  {C.}~\bibnamefont {Felser}}, \bibinfo {author} {\bibfnamefont
  {E.}~\bibnamefont {Hassinger}}, \ and\ \bibinfo {author} {\bibfnamefont
  {B.}~\bibnamefont {Yan}},\ }\href {\doibase 10.1038/ncomms11615} {\bibfield
  {journal} {\bibinfo  {journal} {Nature Communications}\ }\textbf {\bibinfo
  {volume} {7}} (\bibinfo {year} {2016}),\ 10.1038/ncomms11615}\BibitemShut
  {NoStop}%
\bibitem [{\citenamefont {Klotz}\ \emph {et~al.}(2016)\citenamefont {Klotz},
  \citenamefont {Wu}, \citenamefont {Shekhar}, \citenamefont {Sun},
  \citenamefont {Schmidt}, \citenamefont {Nicklas}, \citenamefont {Baenitz},
  \citenamefont {Uhlarz}, \citenamefont {Wosnitza}, \citenamefont {Felser},\
  and\ \citenamefont {Yan}}]{Klotz:2016}%
  \BibitemOpen
  \bibfield  {author} {\bibinfo {author} {\bibfnamefont {J.}~\bibnamefont
  {Klotz}}, \bibinfo {author} {\bibfnamefont {S.-C.}\ \bibnamefont {Wu}},
  \bibinfo {author} {\bibfnamefont {C.}~\bibnamefont {Shekhar}}, \bibinfo
  {author} {\bibfnamefont {Y.}~\bibnamefont {Sun}}, \bibinfo {author}
  {\bibfnamefont {M.}~\bibnamefont {Schmidt}}, \bibinfo {author} {\bibfnamefont
  {M.}~\bibnamefont {Nicklas}}, \bibinfo {author} {\bibfnamefont
  {M.}~\bibnamefont {Baenitz}}, \bibinfo {author} {\bibfnamefont
  {M.}~\bibnamefont {Uhlarz}}, \bibinfo {author} {\bibfnamefont
  {J.}~\bibnamefont {Wosnitza}}, \bibinfo {author} {\bibfnamefont
  {C.}~\bibnamefont {Felser}}, \ and\ \bibinfo {author} {\bibfnamefont
  {B.}~\bibnamefont {Yan}},\ }\href {\doibase 10.1103/PhysRevB.93.121105}
  {\bibfield  {journal} {\bibinfo  {journal} {Phys. Rev. B}\ }\textbf {\bibinfo
  {volume} {93}},\ \bibinfo {pages} {121105} (\bibinfo {year}
  {2016})}\BibitemShut {NoStop}%
\bibitem [{\citenamefont {Shekhar}\ \emph {et~al.}(2015)\citenamefont
  {Shekhar}, \citenamefont {Nayak}, \citenamefont {Sun}, \citenamefont
  {Schmidt}, \citenamefont {Nicklas}, \citenamefont {Leermakers}, \citenamefont
  {Zeitler}, \citenamefont {Skourski}, \citenamefont {Wosnitza}, \citenamefont
  {Liu}, \citenamefont {Chen}, \citenamefont {Schnelle}, \citenamefont
  {Borrmann}, \citenamefont {Grin}, \citenamefont {Felser},\ and\ \citenamefont
  {Yan}}]{Shekhar2015}%
  \BibitemOpen
  \bibfield  {author} {\bibinfo {author} {\bibfnamefont {C.}~\bibnamefont
  {Shekhar}}, \bibinfo {author} {\bibfnamefont {A.~K.}\ \bibnamefont {Nayak}},
  \bibinfo {author} {\bibfnamefont {Y.}~\bibnamefont {Sun}}, \bibinfo {author}
  {\bibfnamefont {M.}~\bibnamefont {Schmidt}}, \bibinfo {author} {\bibfnamefont
  {M.}~\bibnamefont {Nicklas}}, \bibinfo {author} {\bibfnamefont
  {I.}~\bibnamefont {Leermakers}}, \bibinfo {author} {\bibfnamefont
  {U.}~\bibnamefont {Zeitler}}, \bibinfo {author} {\bibfnamefont
  {Y.}~\bibnamefont {Skourski}}, \bibinfo {author} {\bibfnamefont
  {J.}~\bibnamefont {Wosnitza}}, \bibinfo {author} {\bibfnamefont
  {Z.}~\bibnamefont {Liu}}, \bibinfo {author} {\bibfnamefont {Y.}~\bibnamefont
  {Chen}}, \bibinfo {author} {\bibfnamefont {W.}~\bibnamefont {Schnelle}},
  \bibinfo {author} {\bibfnamefont {H.}~\bibnamefont {Borrmann}}, \bibinfo
  {author} {\bibfnamefont {Y.}~\bibnamefont {Grin}}, \bibinfo {author}
  {\bibfnamefont {C.}~\bibnamefont {Felser}}, \ and\ \bibinfo {author}
  {\bibfnamefont {B.}~\bibnamefont {Yan}},\ }\href {\doibase 10.1038/NPHYS3372}
  {\bibfield  {journal} {\bibinfo  {journal} {Nat. Phys.}\ } (\bibinfo {year}
  {2015}),\ 10.1038/NPHYS3372}\BibitemShut {NoStop}%
\bibitem [{\citenamefont {Moll}\ \emph {et~al.}(2016)\citenamefont {Moll},
  \citenamefont {Potter}, \citenamefont {Nair}, \citenamefont {Ramshaw},
  \citenamefont {Modic}, \citenamefont {Riggs}, \citenamefont {Zeng},
  \citenamefont {Ghimire}, \citenamefont {Bauer}, \citenamefont {Kealhofer},
  \citenamefont {Ronning},\ and\ \citenamefont {Analytis}}]{Moll2016}%
  \BibitemOpen
  \bibfield  {author} {\bibinfo {author} {\bibfnamefont {P.~J.~W.}\
  \bibnamefont {Moll}}, \bibinfo {author} {\bibfnamefont {A.~C.}\ \bibnamefont
  {Potter}}, \bibinfo {author} {\bibfnamefont {N.~L.}\ \bibnamefont {Nair}},
  \bibinfo {author} {\bibfnamefont {B.~J.}\ \bibnamefont {Ramshaw}}, \bibinfo
  {author} {\bibfnamefont {K.~A.}\ \bibnamefont {Modic}}, \bibinfo {author}
  {\bibfnamefont {S.}~\bibnamefont {Riggs}}, \bibinfo {author} {\bibfnamefont
  {B.}~\bibnamefont {Zeng}}, \bibinfo {author} {\bibfnamefont {N.~J.}\
  \bibnamefont {Ghimire}}, \bibinfo {author} {\bibfnamefont {E.~D.}\
  \bibnamefont {Bauer}}, \bibinfo {author} {\bibfnamefont {R.}~\bibnamefont
  {Kealhofer}}, \bibinfo {author} {\bibfnamefont {F.}~\bibnamefont {Ronning}},
  \ and\ \bibinfo {author} {\bibfnamefont {J.~G.}\ \bibnamefont {Analytis}},\
  }\href {\doibase 10.1038/ncomms12492} {\bibfield  {journal} {\bibinfo
  {journal} {Nature Communications}\ }\textbf {\bibinfo {volume} {7}},\
  \bibinfo {pages} {12492} (\bibinfo {year} {2016})}\BibitemShut {NoStop}%
\bibitem [{\citenamefont {Chen}\ and\ \citenamefont {Franz}(2016)}]{chen:2016}%
  \BibitemOpen
  \bibfield  {author} {\bibinfo {author} {\bibfnamefont {A.}~\bibnamefont
  {Chen}}\ and\ \bibinfo {author} {\bibfnamefont {M.}~\bibnamefont {Franz}},\
  }\href {\doibase 10.1103/PhysRevB.93.201105} {\bibfield  {journal} {\bibinfo
  {journal} {Phys. Rev. B}\ }\textbf {\bibinfo {volume} {93}},\ \bibinfo
  {pages} {201105} (\bibinfo {year} {2016})}\BibitemShut {NoStop}%
\bibitem [{\citenamefont {Lucas}\ \emph {et~al.}(2016)\citenamefont {Lucas},
  \citenamefont {Davison},\ and\ \citenamefont {Sachdev}}]{Lucas:2016}%
  \BibitemOpen
  \bibfield  {author} {\bibinfo {author} {\bibfnamefont {A.}~\bibnamefont
  {Lucas}}, \bibinfo {author} {\bibfnamefont {R.~A.}\ \bibnamefont {Davison}},
  \ and\ \bibinfo {author} {\bibfnamefont {S.}~\bibnamefont {Sachdev}},\
  }\href@noop {} {\bibfield  {journal} {\bibinfo  {journal} {Proceedings of the
  National Academy of Sciences}\ ,\ \bibinfo {pages} {201608881}} (\bibinfo
  {year} {2016})}\BibitemShut {NoStop}%
\bibitem [{\citenamefont {Ramshaw}\ \emph {et~al.}(2017)\citenamefont
  {Ramshaw}, \citenamefont {Modic}, \citenamefont {Shekhter}, \citenamefont
  {Zhang}, \citenamefont {Kim}, \citenamefont {Moll}, \citenamefont {Bachmann},
  \citenamefont {Chan}, \citenamefont {Betts}, \citenamefont {Balakirev},
  \citenamefont {Migliori}, \citenamefont {Ghimire}, \citenamefont {Bauer},
  \citenamefont {Ronning},\ and\ \citenamefont {McDonald}}]{Ramshaw:2017}%
  \BibitemOpen
  \bibfield  {author} {\bibinfo {author} {\bibfnamefont {B.~J.}\ \bibnamefont
  {Ramshaw}}, \bibinfo {author} {\bibfnamefont {K.~A.}\ \bibnamefont {Modic}},
  \bibinfo {author} {\bibfnamefont {A.}~\bibnamefont {Shekhter}}, \bibinfo
  {author} {\bibfnamefont {Y.}~\bibnamefont {Zhang}}, \bibinfo {author}
  {\bibfnamefont {E.-A.}\ \bibnamefont {Kim}}, \bibinfo {author} {\bibfnamefont
  {P.~J.~W.}\ \bibnamefont {Moll}}, \bibinfo {author} {\bibfnamefont
  {M.}~\bibnamefont {Bachmann}}, \bibinfo {author} {\bibfnamefont {M.~K.}\
  \bibnamefont {Chan}}, \bibinfo {author} {\bibfnamefont {J.~B.}\ \bibnamefont
  {Betts}}, \bibinfo {author} {\bibfnamefont {F.}~\bibnamefont {Balakirev}},
  \bibinfo {author} {\bibfnamefont {A.}~\bibnamefont {Migliori}}, \bibinfo
  {author} {\bibfnamefont {N.~J.}\ \bibnamefont {Ghimire}}, \bibinfo {author}
  {\bibfnamefont {E.~D.}\ \bibnamefont {Bauer}}, \bibinfo {author}
  {\bibfnamefont {F.}~\bibnamefont {Ronning}}, \ and\ \bibinfo {author}
  {\bibfnamefont {R.~D.}\ \bibnamefont {McDonald}},\ }\href@noop {} {\enquote
  {\bibinfo {title} {Unmasking weyl fermions using extreme magnetic fields},}\
  } (\bibinfo {year} {2017}),\ \Eprint {http://arxiv.org/abs/arXiv:1704.06944}
  {arXiv:1704.06944} \BibitemShut {NoStop}%
\bibitem [{\citenamefont {Gooth}\ \emph {et~al.}(2017)\citenamefont {Gooth},
  \citenamefont {Niemann}, \citenamefont {Meng}, \citenamefont {Grushin},
  \citenamefont {Landsteiner}, \citenamefont {Gotsmann}, \citenamefont
  {Menges}, \citenamefont {Schmidt}, \citenamefont {Shekhar}, \citenamefont
  {S{\"{u}}{\ss}}, \citenamefont {H\"uhne}, \citenamefont {Rellinghaus},
  \citenamefont {Felser}, \citenamefont {Yan},\ and\ \citenamefont
  {Nielsch}}]{Gooth:2017}%
  \BibitemOpen
  \bibfield  {author} {\bibinfo {author} {\bibfnamefont {J.}~\bibnamefont
  {Gooth}}, \bibinfo {author} {\bibfnamefont {A.~C.}\ \bibnamefont {Niemann}},
  \bibinfo {author} {\bibfnamefont {T.}~\bibnamefont {Meng}}, \bibinfo {author}
  {\bibfnamefont {A.~G.}\ \bibnamefont {Grushin}}, \bibinfo {author}
  {\bibfnamefont {K.}~\bibnamefont {Landsteiner}}, \bibinfo {author}
  {\bibfnamefont {B.}~\bibnamefont {Gotsmann}}, \bibinfo {author}
  {\bibfnamefont {F.}~\bibnamefont {Menges}}, \bibinfo {author} {\bibfnamefont
  {M.}~\bibnamefont {Schmidt}}, \bibinfo {author} {\bibfnamefont
  {C.}~\bibnamefont {Shekhar}}, \bibinfo {author} {\bibfnamefont
  {V.}~\bibnamefont {S{\"{u}}{\ss}}}, \bibinfo {author} {\bibfnamefont
  {R.}~\bibnamefont {H\"uhne}}, \bibinfo {author} {\bibfnamefont
  {B.}~\bibnamefont {Rellinghaus}}, \bibinfo {author} {\bibfnamefont
  {C.}~\bibnamefont {Felser}}, \bibinfo {author} {\bibfnamefont
  {B.}~\bibnamefont {Yan}}, \ and\ \bibinfo {author} {\bibfnamefont
  {K.}~\bibnamefont {Nielsch}},\ }\href@noop {} {\bibfield  {journal} {\bibinfo
   {journal} {Nature}\ }\textbf {\bibinfo {volume} {547}},\ \bibinfo {pages}
  {324} (\bibinfo {year} {2017})}\BibitemShut {NoStop}%
\bibitem [{\citenamefont {Wei}\ \emph {et~al.}(2012)\citenamefont {Wei},
  \citenamefont {Chao},\ and\ \citenamefont {Aji}}]{Wei:2012}%
  \BibitemOpen
  \bibfield  {author} {\bibinfo {author} {\bibfnamefont {H.}~\bibnamefont
  {Wei}}, \bibinfo {author} {\bibfnamefont {S.-P.}\ \bibnamefont {Chao}}, \
  and\ \bibinfo {author} {\bibfnamefont {V.}~\bibnamefont {Aji}},\ }\href
  {\doibase 10.1103/PhysRevLett.109.196403} {\bibfield  {journal} {\bibinfo
  {journal} {Phys. Rev. Lett.}\ }\textbf {\bibinfo {volume} {109}},\ \bibinfo
  {pages} {196403} (\bibinfo {year} {2012})}\BibitemShut {NoStop}%
\bibitem [{\citenamefont {Cho}\ \emph {et~al.}(2012)\citenamefont {Cho},
  \citenamefont {Bardarson}, \citenamefont {Lu},\ and\ \citenamefont
  {Moore}}]{Cho:2012}%
  \BibitemOpen
  \bibfield  {author} {\bibinfo {author} {\bibfnamefont {G.~Y.}\ \bibnamefont
  {Cho}}, \bibinfo {author} {\bibfnamefont {J.~H.}\ \bibnamefont {Bardarson}},
  \bibinfo {author} {\bibfnamefont {Y.-M.}\ \bibnamefont {Lu}}, \ and\ \bibinfo
  {author} {\bibfnamefont {J.~E.}\ \bibnamefont {Moore}},\ }\href {\doibase
  10.1103/PhysRevB.86.214514} {\bibfield  {journal} {\bibinfo  {journal} {Phys.
  Rev. B}\ }\textbf {\bibinfo {volume} {86}},\ \bibinfo {pages} {214514}
  (\bibinfo {year} {2012})}\BibitemShut {NoStop}%
\bibitem [{\citenamefont {Meng}\ and\ \citenamefont
  {Balents}(2012)}]{Meng:2012}%
  \BibitemOpen
  \bibfield  {author} {\bibinfo {author} {\bibfnamefont {T.}~\bibnamefont
  {Meng}}\ and\ \bibinfo {author} {\bibfnamefont {L.}~\bibnamefont {Balents}},\
  }\href {\doibase 10.1103/PhysRevB.86.054504} {\bibfield  {journal} {\bibinfo
  {journal} {Phys. Rev. B}\ }\textbf {\bibinfo {volume} {86}},\ \bibinfo
  {pages} {054504} (\bibinfo {year} {2012})}\BibitemShut {NoStop}%
\bibitem [{\citenamefont {Zhang}\ and\ \citenamefont
  {Nagaosa}(2017)}]{Zhang:2017b}%
  \BibitemOpen
  \bibfield  {author} {\bibinfo {author} {\bibfnamefont {X.-X.}\ \bibnamefont
  {Zhang}}\ and\ \bibinfo {author} {\bibfnamefont {N.}~\bibnamefont
  {Nagaosa}},\ }\href {\doibase 10.1103/PhysRevB.95.205143} {\bibfield
  {journal} {\bibinfo  {journal} {Phys. Rev. B}\ }\textbf {\bibinfo {volume}
  {95}},\ \bibinfo {pages} {205143} (\bibinfo {year} {2017})}\BibitemShut
  {NoStop}%
\bibitem [{\citenamefont {Herring}(1937)}]{herring:1937}%
  \BibitemOpen
  \bibfield  {author} {\bibinfo {author} {\bibfnamefont {C.}~\bibnamefont
  {Herring}},\ }\href {\doibase 10.1103/PhysRev.52.365} {\bibfield  {journal}
  {\bibinfo  {journal} {Phys. Rev.}\ }\textbf {\bibinfo {volume} {52}},\
  \bibinfo {pages} {365} (\bibinfo {year} {1937})}\BibitemShut {NoStop}%
\bibitem [{\citenamefont {Vafek}\ and\ \citenamefont
  {Vishwanath}(2014)}]{Vafek:2014}%
  \BibitemOpen
  \bibfield  {author} {\bibinfo {author} {\bibfnamefont {O.}~\bibnamefont
  {Vafek}}\ and\ \bibinfo {author} {\bibfnamefont {A.}~\bibnamefont
  {Vishwanath}},\ }\href@noop {} {\bibfield  {journal} {\bibinfo  {journal}
  {Annu. Rev. Condens. Matter Phys.}\ }\textbf {\bibinfo {volume} {5}},\
  \bibinfo {pages} {83} (\bibinfo {year} {2014})}\BibitemShut {NoStop}%
\end{thebibliography}%

\end{document}